\documentclass[prl,aps,twocolumn,superscriptaddress,]{revtex4}
\usepackage{amsfonts}
\usepackage{dcolumn}
\usepackage{bm}
\usepackage{tikz}
\usepackage[colorlinks=true,citecolor=blue,urlcolor=blue]{hyperref}
\usepackage{amsmath,amsfonts,amssymb,times,natbib}
\usepackage[standard]{ntheorem}

\begin{document}
\renewcommand{\figurename}{Fig.}

\title{Zonal reconstruction of photonic wavefunction via momentum weak measurement}

\author{Mu Yang}
\affiliation{CAS Key Laboratory of Quantum Information, University of Science and Technology of China, Hefei 230026, People's Republic of China}
\affiliation{CAS Center For Excellence in Quantum Information and Quantum Physics, University of Science and Technology of China, Hefei 230026, People's Republic of China}

\author{Ya Xiao} 
\affiliation{Department of Physics, Ocean University of China, Qingdao 266100, People's Republic of China}

\author{Yi-Wei Liao}
\affiliation{CAS Key Laboratory of Quantum Information, University of Science and Technology of China, Hefei 230026, People's Republic of China}
\affiliation{CAS Center For Excellence in Quantum Information and Quantum Physics, University of Science and Technology of China, Hefei 230026, People's Republic of China}

\author{Zheng-Hao Liu}
\affiliation{CAS Key Laboratory of Quantum Information, University of Science and Technology of China, Hefei 230026, People's Republic of China}
\affiliation{CAS Center For Excellence in Quantum Information and Quantum Physics, University of Science and Technology of China, Hefei 230026, People's Republic of China}

\author{Xiao-Ye Xu}
\affiliation{CAS Key Laboratory of Quantum Information, University of Science and Technology of China, Hefei 230026, People's Republic of China}
\affiliation{CAS Center For Excellence in Quantum Information and Quantum Physics, University of Science and Technology of China, Hefei 230026, People's Republic of China}

\author{Jin-Shi Xu}
\email{jsxu@ustc.edu.cn}
\affiliation{CAS Key Laboratory of Quantum Information, University of Science and Technology of China, Hefei 230026, People's Republic of China}
\affiliation{CAS Center For Excellence in Quantum Information and Quantum Physics, University of Science and Technology of China, Hefei 230026, People's Republic of China}

\author{Chuan-Feng Li}
\email{cfli@ustc.edu.cn}
\affiliation{CAS Key Laboratory of Quantum Information, University of Science and Technology of China, Hefei 230026, People's Republic of China}
\affiliation{CAS Center For Excellence in Quantum Information and Quantum Physics, University of Science and Technology of China, Hefei 230026, People's Republic of China}

\author{Guang-Can Guo}
\affiliation{CAS Key Laboratory of Quantum Information, University of Science and Technology of China, Hefei 230026, People's Republic of China}
\affiliation{CAS Center For Excellence in Quantum Information and Quantum Physics, University of Science and Technology of China, Hefei 230026, People's Republic of China}

\date{\today}

\begin{abstract}
{
	Direct measurement of wave functions has attracted great interests and many different methods have been developed. However, the precision of current techniques is limited by the use of Fourier transform lenses. 
	These measurements require to shear cut the part of particles with momentum $P=0$, which greatly restricts the efficiency and application of the approaches. Here, we propose and experimentally demonstrate a method to directly measure two-dimensional photonic wave functions by combining the momentum weak measurement technology and the zonal wavefront restoration algorithm. 
	Both the Gaussian and Laguerre-Gaussian wave functions are experimentally well reconstructed. Our method avoids using the Fourier lens and post selection on the momentum $P=0$. We further apply it to measure wavefronts with ultra-high spatial frequency, which is difficult for traditional Shack-Hartmann wavefront sensing technologies. 
	Our work extends the ability of quantum weak measurement and would be useful for wavefront sensing.
}
\end{abstract}
	\maketitle
	
	\section{Introduction}
	
	The construction of the wave function of a quantum system is of fundamental and practical importance. Quantum state tomography \cite{breitenbach1997measurement,smithey1993measurement} is an indirect method to establish the quantum states, which requires a large set of strong measurements and post-processing of these information. 
	In 2011, Lundeen \emph{et al}. introduced a direct measurement of quantum states via weak measurement~\cite{lundeen2011direct}, which reduced the complexity in characterizing a quantum system. 
	Thereafter great interests have been attracted to extend the application to different types of quantum system, such as mixed~\cite{lundeen2012procedure} and high-dimensional~\cite{mirhosseini2014compressive,malik2014direct} states. In the technical aspect, the proposal of direct measurement via scan-free measurement~\cite{shi2015scan} and strong measurement~\cite{zhang2018direct} further improved the technique of direct wave function construction. 
	
	In general, the direct measurement of photonic wave functions consists the Fourier transform lens to achieve transformation of coordinate space and position space, in order to shear cut the part of particles with the momentum $P=0$. The extraction on the part of $P=0$ increases the experimental complexity and it usually causes the loss of photons. In addition, the photonic wave function that is outside from $P=0$ (higher modes) is always hard to reconstruct.
	On the other hand, the experiments~\cite{lundeen2011direct, lundeen2012procedure, mirhosseini2014compressive, malik2014direct} considering the weak measurement of positions followed by a strong measurement of the momentum, often require scanning in the coordinate space, which is time consuming and would be hard to realize in characterizing high dimensional quantum systems.
	
	Here, we demonstrate a Lens-less and scan-free direct measurement of photonic two-dimensional transverse wave functions. We obtain the momentum information directly via weak measurement~\cite{kocsis2011observing,xiao2017experimental} rather than shear cut the photons at the momentum $P=0$.
	Combining with the zonal wavefront reconstruction algorithm~\cite{Regional}, the photonic two-dimensional transverse wave function can be established. Both the Gaussian and Laguerre-Gaussian wave functions~\cite{Allen} are experimentally well reconstructed. 
	
	Naturally, the wave function reconstruction is closely related to the wavefront sensing, which is widely used in astronomy~\cite{roddier1999adaptive,lofdahl1994wavefront}, medicine~\cite{liang1994objective,prieto2000analysis} and laser technology~\cite{roorda2002adaptive}. The traditional Shack-Hartmann wavefront sensor (SHWFS)~\cite{Hartmann, platt1971lenticular} has the difficulty in realizing accurate sampling of wavefront slopes for the limitation of lens arrays, which results in a low spatial resolution of reconstructed wavefronts.
	Our work provides a new tack for high precision wave-front slopes sampling by detecting momentum weak values, and achieves pixel-level wavefront reconstruction, which is of great significance for wavefront sensing. We further experimentally reconstruct wavefronts which are diffused by a strong scattering medium. The results show that this method has advantages in wavefront sensing with a high spatial frequency.

\begin{figure*}[ht]
	\centering
	\includegraphics[width=0.85\linewidth]{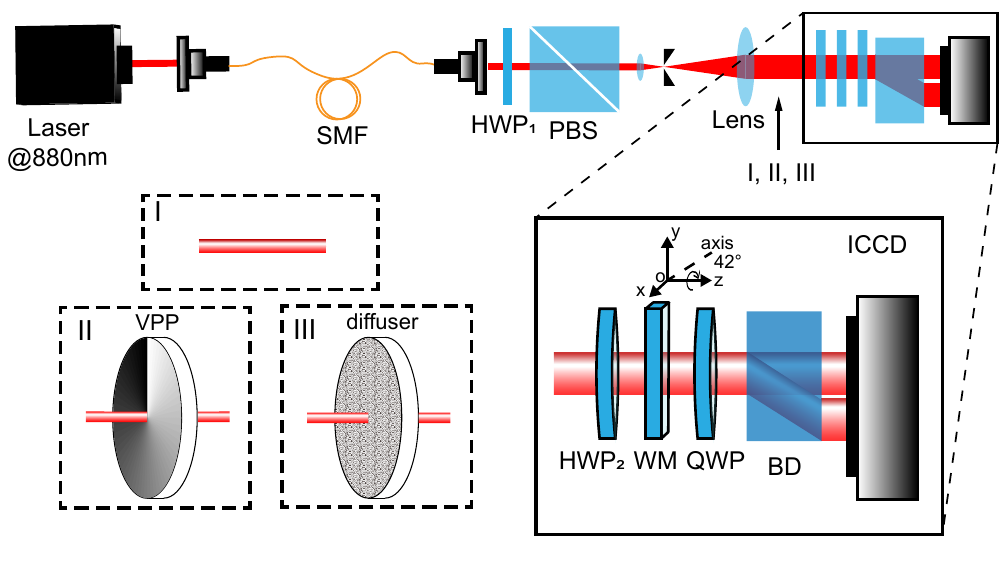}
	\caption{\textbf{Experimental setup.} The laser with a central wavelength of 880 nm is coupled to a single mode fiber (SMF). A half-wave plate (HWP$_1$) and a polarization beam splitter (PBS) are used to rotate the beam's polarization. The wavefront is shaped into a Gaussian mode with a spatial filter, which is consisted of two lenses and a pinhole between them. After passing through another HWP$_2$, the light beam is weakly measured by a thin birefringent crystal (WM) with the axis set to be $42^\circ$ in the y-z plane, which can be rotated along the z axis to weakly couple the polarization and momentums in different dimensions. A quarter-wave plate (QWP) and a beam displacer (BD) are used to project polarization states. The light beams are finally detected in an intensified charge coupled device (ICCD) camera. 
	Three different input modes are used: (I) the free Gaussian mode; (II) the Laguerre-Gaussian mode by passing the light through a vortex phase plate (VPP) with a $2\pi$ phase rotation; (III) the scattered mode with the light beam passing through a diffuser (600 grits).
	}
	\label{setup}
\end{figure*}

\begin{figure}[t]
	\centering
	\includegraphics[width=1 \linewidth]{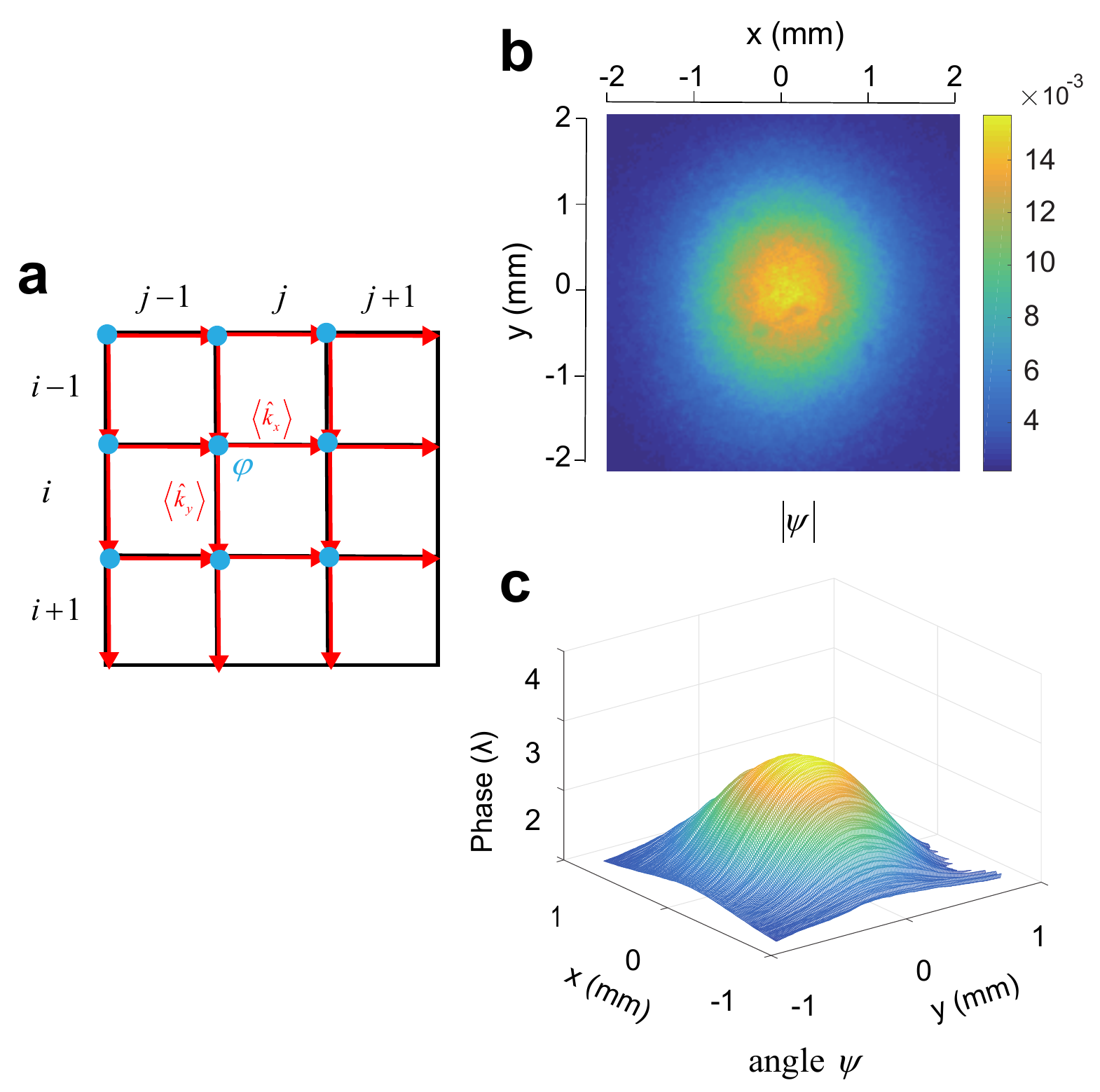}
	\caption{ \textbf{Experimental results of the Gaussian mode}. 
	\textbf{a.} Hudgin model.
	\textbf{b.} The Gaussian beam intensity distribution on the ICCD camera. 
	\textbf{c.} The reconstructed Gaussian phase distribution.}
	\label{Gauss}
\end{figure}

	\section{Theoretical framework}
	
	In 1988, Aharonov, Albert and Vaidman firstly proposed the concept of weak measurements~\cite{aharonov1988result}, which required very weak interaction between the system and a pointer system. 
	Usually, weak measurements of wave functions rest on the sequential measurements of conjugate observables. The first measurement of the momentum along the direction \textbf{r} is weak enough followed by a strong measurement on the position (also known as "post selection"). Assuming that the observable of the system to be measured is the momentum $\hat{P}_r=-i\hbar\frac{\partial }{\partial \vec{r}}$, the initial state is $|\psi\rangle$, and the "post-selected" state is the coordinate $|r\rangle$, the momentum weak value (taking $\hbar=1$) is given as,
	\begin{equation}
	\langle\hat{k}_{r}\rangle=\frac{\langle r|\hat{P}_{r}|\psi\rangle}{\langle r|\psi\rangle}.
	\end{equation}
	If the direction $\vec{r}$ is chosen as $\vec{x}$ ($\vec{y}$), we can get the horizontal (vertical) momentum of particles $\langle\hat{k}_{x(y)}\rangle$. 
	According to the momentum distribution, we can reconstruct the phase of the wave function,
	\begin{equation}
	\varphi= \int \frac{\langle\hat{k}_{x}\rangle}{|\textbf{k}|} dx + \frac{\langle\hat{k}_{y}\rangle}{|\textbf{k}|} dy,
	\end{equation}
	where $|\textbf{k}|=\frac{2\pi}{\lambda}$, of which $\lambda$ represents the wavelength. The transverse wave function is then given by $|\psi\rangle=|\psi|e^{-i\varphi}$,
    where $|\psi|$ is the amplitude and can be obtained via the projection measurement. See more theoretical details in the Methods. 
	
	\begin{figure*}[ht!]
		\centering
		\includegraphics[width=0.92\textwidth]{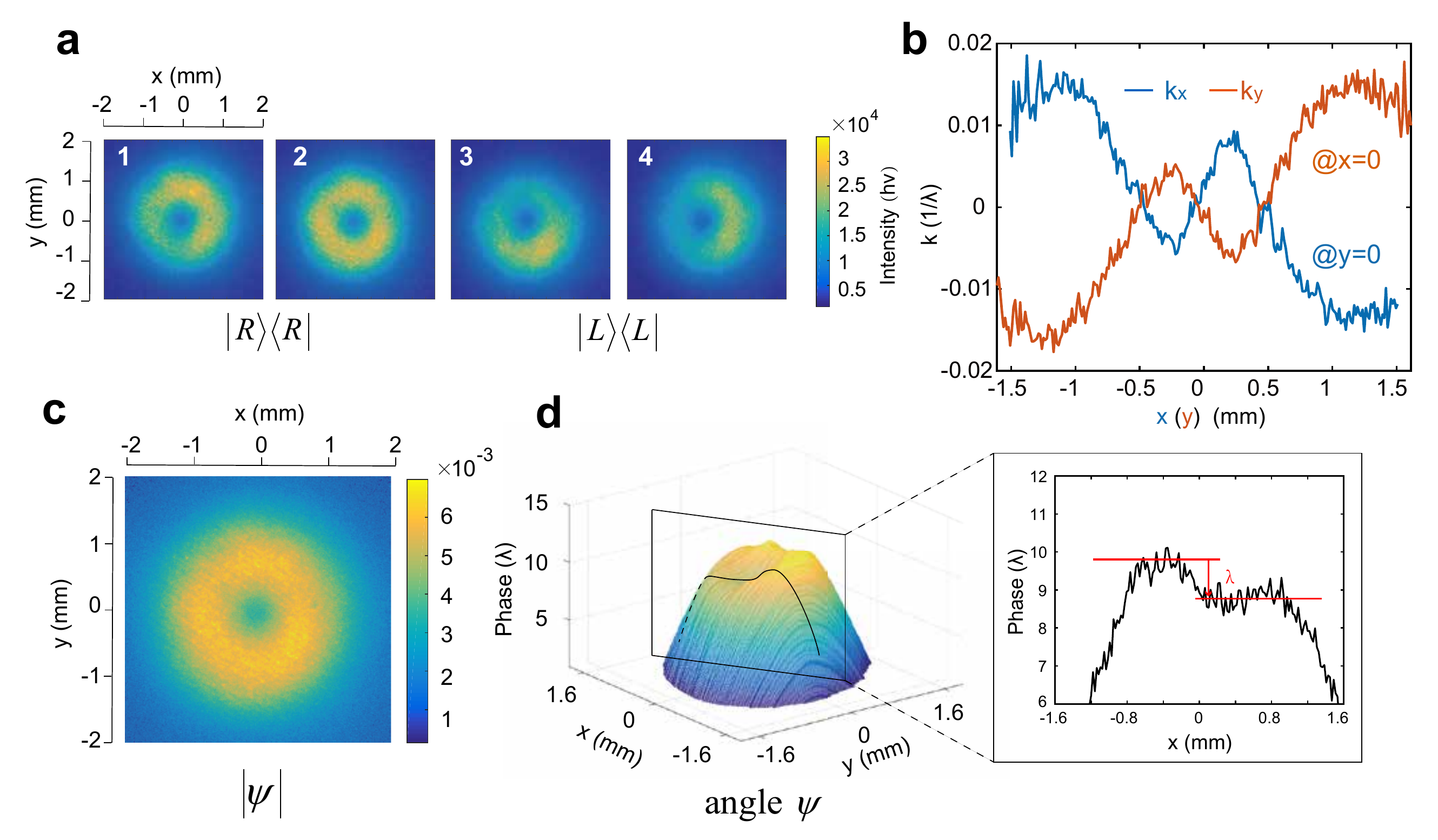}
		\caption{ \textbf{Experimental results of the Laguerre-Gaussian mode}. \textbf{a.} When the optical axis of the weak measurement crystal is set in the x-z  (y-z) plane, the intensity distributions of the light beam projected to the right/left-handed circular polarization ($|R\rangle \langle R|/|L\rangle \langle L|$) are represented in 1 and 3 (2 and 4), respectively.  \textbf{b.} Transverse momentum distributions in the x/y direction at y/x=0. \textbf{c.} The amplitude of the Laguerre-Gaussian wave function. \textbf{d.} The phase distribution of the Laguerre-Gaussian wave function.  The inset shows the magnified case in the black box, which represents the phase distribution in the x direction at y=-0.4 mm. There is a $\lambda$ phase difference in the x direction.
		}
		\label{LG}
	\end{figure*}
	
	\begin{figure*}[ht!]
		\centering
		\includegraphics[width=0.92 \linewidth]{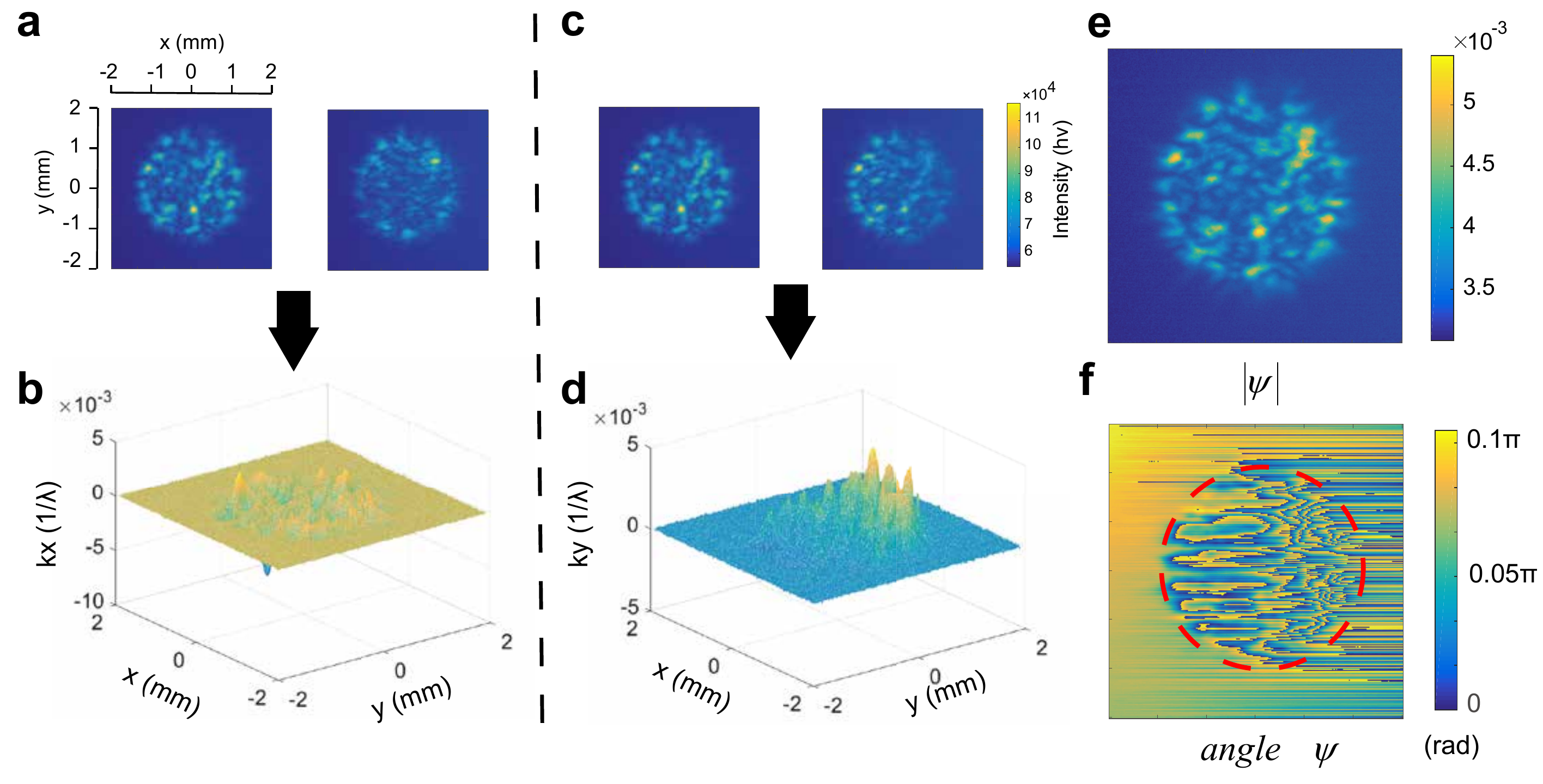}
		\caption{\textbf{Experimental results of the scattered light beam. } \textbf{a}. The intensity distributions projected to the right-handed circular polarization (left figure) and the left-handed circular polarization (right figure), when the crystal optical axis is set in the x-z plane. \textbf{b}. The transverse momentum for $\langle k_x \rangle$. \textbf{c}. When the crystal optical axis is set in the y-z plane, the intensity distributions projected to the right-handed circular polarization (left figure) and the left-handed circular polarization (right figure). \textbf{d}. The transverse momentum distribution for $\langle k_y \rangle$. \textbf{e}. The amplitude of wave function. \textbf{f}. The phase distribution of wave function denoted in the red circle.
		}
		\label{MG}
	\end{figure*}
	
	By encoding the polarization of photons as the pointer system, the photon's momentum weak value can be detected by using birefringent crystals \cite{xiao2017experimental}, in which no Fourier lenses and shear cut photons at $P=0$ are needed. 
	In addition, the post selection on $\vec{x}$ ($\vec{y}$) can be performed on all position states at the same time through the use of an array photodetector. Therefore the momentum weak value $\langle\hat{k}_{x(y)}\rangle$ at all positions can be measured simultaneously, which does not require scanning.

	\section{Experimental results}
	The experimental setup is shown in Fig. \ref{setup}. An infrared laser with a central wavelength of 880 nm is used and then is coupled into a single mode fiber (SMF). A half wave plate (HWP$_1$) and a polarization beam splitter (PBS) are used to set the polarization. A spatial filter consists two lenses with focal lengths of 50 mm and 100 mm, respectively, and a pinhole with 50 um diameter placing between them. The beam diameter is expanded to about 1 mm after the spatial filter. 
	Here three optical channels are used: the Guassian light beam (I) passes directly; (II) passes though a vortex phase plate (VPP)~\cite{Genevet} to prepared the Laguerre-Gaussian mode with the orbital angular momenta of  l=1; (III) passes though a diffuser (600 grits) to prepared a scattered mode.	
	Another HWP$_2$ set at $22.5^{\circ}$ is used to rotate the polarization into a superposition state $1/\sqrt{2}(|H\rangle+|V\rangle)$, where $|H(V)\rangle$ represents the horizontal (vertical) polarization. The light beam further passes through a thin calcite with the thickness of 0.7 mm. To measure the momentum weak values along the x axis ($\langle k_x \rangle $), the optical axis is set in the x-z plane, which is oriented at $42^\circ$ with respect to the z axis. To measure the momentum weak values along the y axis ($\langle k_y \rangle $), the calcite is rotated by $90^\circ$ to let the optic axis in the y-z plane. A relative phase is introduced between the vertical and horizontal polarization states, described as $|H\rangle+|V\rangle \to |H\rangle+ e^{i\zeta \langle k_{x(y)} \rangle/k}|V\rangle$, where the dimensionless coupling strength $\zeta$ approximates 336~\cite{xiao2017experimental}. 
	
	A quarter-wave plate (QWP) with the axis set to $45^{\circ}$ and a beam displacer (BD) are used to project the final polarization states into the right/left-hand circular basis ($|R\rangle=\frac{\sqrt{2}}{2}(|H\rangle-i|V\rangle)/ |L\rangle=\frac{\sqrt{2}}{2}(|H\rangle+i|V\rangle)$). The corresponding beam intensities $I^{R}_{x(y)}$ and $I^{L}_{x(y)}$ are detected by an intensified charge coupled device (ICCD) camera, which consists of 1024$\times$1024 pixels and the length of each square pixel is $\Delta=13\ \mu m$.
	The wave function phase gradient can be measured to be
	\begin{equation}\label{kxy}
	\langle\hat{k}_{x(y)}\rangle/|\textbf{k}|=\dfrac{1}{\zeta}{\rm arcsin}(\dfrac{I^{R}_{x(y)}-I^{L}_{x(y)}}{I^{R}_{x(y)}+I^{L}_{x(y)}}).
	\end{equation}
     For each pixel $(i, j)$, the discrete wavefront phase distribution  $\varphi$ can be obtained as (also known as Hudgin model~\cite{hudgin1977wave}):
	\begin{eqnarray}
	\label{k1}
	\langle\hat{k}_{x}(i,j)\rangle/|\textbf{k}|& = &\dfrac{\varphi_{i+1,j}-\varphi_{i,j}}{\Delta},\\
	\langle\hat{k}_{y}(i,j)\rangle/|\textbf{k}|& = &\dfrac{\varphi_{i,j+1}-\varphi_{i,j}}{\Delta}.
	\label{k2}
	\end{eqnarray}
	According to Eqs. (\ref{k1}) and (\ref{k2}), we can iteratively calculate the wave function phase distribution, which is shown in Fig. \ref{Gauss}a. The black edge grids represent the pixels of ICCD, while the blue dots donate the phase points and the red ``$\to$" represent the directions and positions of momenta.
	The final wave function can be obtained through the zonal wavefront reconstruction algorithm as $|\phi\rangle=\sqrt{I_{a.u.}}e^{-i\varphi}$, where $I_{a.u.}$ is the normalization beam intensity distribution directly measured by the ICCD camera. There is no photon loss in our experiment. The complete wavefront phase can then obtained by the zonal wavefront reconstruction algorithm.
	
	We firstly employ the proposed method to directly measure the Gaussian spatial wave function. The initial intensity distribution of each pixel $I_{Gauss}$ is detected directly by ICCD. We can get normalized light intensity $I_{a.u.}=I_{Gauss}/\sum I_{Gauss}$, which represents  $|\psi|^2$. The corresponding experimental $|\psi|$ is shown in Fig. \ref{Gauss}b. We further directly obtain the phase distribution as shown in Fig. \ref{Gauss}c, which is a typical Gaussian phase distribution.
	
	We further measure the wave function of high order modes. The initial wavefunction is prepared to be the Laguerre-Gaussian mode with the orbital angular momenta of $l=1$. The panels signed with 1 and 3 in Fig. \ref{LG}a represents the intensity distributions with left/right-handed circular polarizations, respectively, when measuring $\langle k_x\rangle$. 
	While, the panes signed with 2 and 4 in Fig. \ref{LG}a represent the intensity distributions with left/right-handed circular polarizations, respectively, when measuring $\langle k_y \rangle$.	
	The brown and blue lines in Fig. \ref{LG}b show the variation of $\langle k_x\rangle$ at y=0 and $\langle k_y\rangle$ at x=0, respectively. When closing to the center, the momentum reversely changes, which results that the photons travel spirally. The momentums along x and y directions distribute symmetrically because the wave function is spatially symmetric.
	The normalized amplitude $|\psi|$ of the Laguerre-Gaussian mode is shown in Fig. \ref{LG}c.
	The phase distribution of the transverse wave functions restored by the momentum information is shown in Fig. \ref{LG}d. The inset shows the phase distribution on the cross section of y=-0.4 mm. It can be clearly observed that the wave function phase of x$>$0 is lagging $2\pi$ behind that of x$<$0, which is consistent with the theoretical Laguerre-Gaussian spatial wave function.



Direct measurement of the photonic wave functions can be naturally applied in wavefronts sensing. Our method can further be used to reconstruct the wavefronts with ultra-high spatial frequency that are always presented after some scattering medium, such as diffusers~\cite{bertolotti2012non,he2013image,yang2018deep}, multi-mode fibers~\cite{redding2013all} and so on. The wavefronts are hard to detect by traditional wavefront sensors, such as SHWFS. 
Based on the two-dimensional momentum weak measurements, we can achieve densely sampling of wavefront slopes, and the wavefront can be restored of pixel-level resolution. Some numerical simulations are shown in the Supplementary Information (SI)~\cite{SI}.

We experimentally reconstruct the wavefront which is scattered by a diffuser (600 grits). When the optical axis of weak measurement crystal is set in x-z plane, the intensity distributions are projected to the right-hand and left-hand circular polarizations, respectively, as shown in Fig. \ref{MG}a right (left) panel is the intensity distribution project to the right (left) - hand circular polarization). Fig. \ref{MG}b shows the momentum distribution $\langle k_x \rangle$. Similarly, when detecting the momentum distribution $\langle k_y \rangle$, the light beam is post-selected on the right (left)-hand circular polarization, of which the results are shown in Fig. \ref{MG}c and d.
The normalized intensity and phase of the scattered light beam are shown in Fig. \ref{MG}e and f, respectively. For the measurement errors of scattering modes, phase distributions are of stripe noise. By modding $0.1\pi$, we still find the contour phase map obviously which is shown in the red circle in Fig. \ref{MG}f.

We further introduce another wavefront restoration algorithm (Zernike mode method) based on weak measurements, which can extract Zernike aberration coefficients \cite{zernike} in different orders. See SI for detailed results~\cite{SI}.

	\section{Conclusion}
	
	We have proposed a lens-less direct measurement of photonic two-dimensional wave functions, by employing momentum weak values. The Gaussian spatial wave function and Laguerre-Gaussian wave function are experimentally reconstructed. For the higher mode, the $2\pi$ phase retardation of photons carrying orbital angular momentum of momenta $l=1$ was observed.
	
	Our method can be further used for wavefront sensing. Especially, it can be applied to reconstruct wavefronts with high spatial frequencies, which can not be well achieved by the traditional methods. The wavefunction diffused by a scattering medium is also experimentally reconstructed.
	
	The combination of quantum weak measurements and classical wavefront restoring algorithms not only provides a scheme for the photonic wave function reconstruction, but also greatly improves the accuracy of wavefront sensing. Our work extends the ability of weak measurement and would be useful for wavefront sensing. By optimizing the algorithm, the accuracy can be further improved.

	\section{Acknowledgements}
	
	This work was supported by the National Key Research and Development Program of China (Grant No. 2016YFA0302700 and 2017YFA0304100), the National Natural Science Foundation of China (Grants No. 61725504, 11774335 and 11821404), the Key Research Program of Frontier Sciences, Chinese Academy of Sciences (CAS) (Grant No. QYZDY-SSW-SLH003), Science Foundation of the CAS (No. ZDRW-XH-2019-1), Anhui Initiative in Quantum Information Technologies (AHY060300 and AHY020100), the Fundamental Research Funds for the Central Universities (Grant No. WK2030380017 and WK2470000026).

   \section{Conflict of Interest}
   The authors declare no competing financial interests.
   
   \section{Keywords}
   weak measurement, momentum weak values, wavefunction direct measurement, zonal high-resolution wavefront sensing, zonal wavefront restoration algorithm



\end{document}


\renewcommand{\thefigure}{S\arabic{figure}}
\renewcommand\theequation{S\arabic{equation}}

\title{Supplementary information: Zonal reconstruction of photonic wavefunction via momentum weak measurement}

\author{Mu Yang}
\affiliation{CAS Key Laboratory of Quantum Information, University of Science and Technology of China, Hefei 230026, People's Republic of China}
\affiliation{CAS Center For Excellence in Quantum Information and Quantum Physics, University of Science and Technology of China, Hefei 230026, People's Republic of China}

\author{Ya Xiao} 
\affiliation{Department of Physics, Ocean University of China, Qingdao 266100, People's Republic of China}

\author{Yi-Wei Liao}
\affiliation{CAS Key Laboratory of Quantum Information, University of Science and Technology of China, Hefei 230026, People's Republic of China}
\affiliation{CAS Center For Excellence in Quantum Information and Quantum Physics, University of Science and Technology of China, Hefei 230026, People's Republic of China}

\author{Zheng-Hao Liu}
\affiliation{CAS Key Laboratory of Quantum Information, University of Science and Technology of China, Hefei 230026, People's Republic of China}
\affiliation{CAS Center For Excellence in Quantum Information and Quantum Physics, University of Science and Technology of China, Hefei 230026, People's Republic of China}

\author{Xiao-Ye Xu}
\affiliation{CAS Key Laboratory of Quantum Information, University of Science and Technology of China, Hefei 230026, People's Republic of China}
\affiliation{CAS Center For Excellence in Quantum Information and Quantum Physics, University of Science and Technology of China, Hefei 230026, People's Republic of China}

\author{Jin-Shi Xu}
\email{jsxu@ustc.edu.cn}
\affiliation{CAS Key Laboratory of Quantum Information, University of Science and Technology of China, Hefei 230026, People's Republic of China}
\affiliation{CAS Center For Excellence in Quantum Information and Quantum Physics, University of Science and Technology of China, Hefei 230026, People's Republic of China}

\author{Chuan-Feng Li}
\email{cfli@ustc.edu.cn}
\affiliation{CAS Key Laboratory of Quantum Information, University of Science and Technology of China, Hefei 230026, People's Republic of China}
\affiliation{CAS Center For Excellence in Quantum Information and Quantum Physics, University of Science and Technology of China, Hefei 230026, People's Republic of China}

\author{Guang-Can Guo}
\affiliation{CAS Key Laboratory of Quantum Information, University of Science and Technology of China, Hefei 230026, People's Republic of China}
\affiliation{CAS Center For Excellence in Quantum Information and Quantum Physics, University of Science and Technology of China, Hefei 230026, People's Republic of China}

\date{\today}
\begin{abstract}
This document provides supplementary material to "{Zonal reconstruction of photonic wavefunction via momentum weak measurement}". 
The simulation results of wavefront reconstruction with different kinds of aberrations are shown in section I. 
In section II, another method of wavefront reconstruction based on the Zernike mode method and numerical simulated results are proposed.
\end{abstract}
	
	\maketitle

	\section{I. theoretical framework}
	
	{\bf The momentum weak value.}
	
	The photons' initial state is expressed as:
	\begin{equation}
	|\psi_{0}\rangle=|H\rangle|\phi_0\rangle,
	\end{equation}
	where $|\phi_0\rangle$ represents transverse wave function, which is the function of spatial position $(x,y)$. $|H\rangle$ represents the horizontal polarization. After the preparation operator $\hat{I}=|D\rangle\langle D|\otimes|\phi\rangle\langle\phi|$ the state changes to, 
	\begin{equation}
	|\psi_{i}\rangle=\hat{I}|\psi_{0}\rangle=|D\rangle|\phi\rangle,
	\end{equation}
	where $|\phi\rangle$ is the transverse wave function to be measured and $|D\rangle=\frac{1}{\sqrt{2}}(|H\rangle+|V\rangle)$ ($|V\rangle$ represents the vertical polarization). 
	Then the light beam passes through the weak measurement system, which couples polarization pointer $|H\rangle\langle H|$ and momentum operator $\hat{P}_{x(y)}$ along the $x$ $(y)$ direction, where $\hat{P}_{x(y)}$=$-i\hbar\frac{\partial}{\partial x(y)}$. The interaction Hamiltonian for weak measurement is given as $\hat{H}_{x(y)}$=$g|H\rangle\langle H|\otimes\hat{P}_{x(y)}$. $g$ is the coupling strength and set to be very small, so the influence of the measurement on the system is very weak. The time evolution of the Hamiltonian can be expressed as
	\begin{eqnarray}
	\hat{U}_{\omega}|\psi_{i}\rangle&=&e^{-i\hat{H}_{x(y)}\Delta t/\hbar}|\psi_{i}\rangle \\
	&\approx&(1-ig\Delta t|H\rangle\langle H|\otimes\hat{k}_{x(y)})|\psi_{i}\rangle.
	\end{eqnarray}
	We can write the state as
	the state of beam becomes
	\begin{eqnarray}
	\label{WM}
	|\psi^{'}_{i}\rangle & = & e^{-i\hat{H}_{x(y)}\Delta t}|D\rangle|\phi\rangle\\
	& = & 1/\sqrt{2}(|H\rangle e^{-ig\Delta t\langle\hat{k}_{x(y)}\rangle}+|V\rangle)|\phi\rangle,
	\end{eqnarray}
	where $\langle\hat{k}_{x(y)}\rangle$=$\langle x(y)|\hat{k}_{x(y)}|\phi\rangle/\langle x(y)|\phi\rangle$ is the so-called weak momentum value. To measure this value, the pointer state (polarization) is post-selected onto the circular basis $|R\rangle\langle R|$ and $|L\rangle\langle L|$ ($|R\rangle=\frac{\sqrt{2}}{2}(|H\rangle-i|V\rangle), |L\rangle=\frac{\sqrt{2}}{2}(|H\rangle+i|V\rangle)$) at each position of $|x(y)\rangle\langle x(y)|$.
	The corresponding wave function $\psi^{R}_{x(y)}$ and $\psi^{L}_{x(y)}$ are
	\begin{eqnarray}
	\psi^{R(L)}_{x(y)} & = &\langle R(L)|\langle x(y)|\psi^{'}_i\rangle\\
	& = & 1/2(e^{-ig\Delta t\langle\hat{k}_{x(y)}\rangle}\pm i) \langle x(y)|\phi\rangle.
	\end{eqnarray}
	The expression of the weak value of momentum is given as
	\begin{equation}
	\langle\hat{k}_{x(y)}\rangle=\dfrac{1}{g\Delta t}{\rm arcsin}(|\psi^{R}_{x(y)}|^{2}-|\psi^{L}_{x(y)}|^{2}),
	\end{equation}
	We set interaction strength to be $\zeta=g\Delta t|\bf{k}|$. \\

	{\bf The reconstruction of wavefunction.}
	
	The wave function phase part $\varphi$ can be expressed as
	\begin{eqnarray}
	\partial \varphi/\partial x(y)& = & \langle\hat{k}_{x(y)}\rangle/|\textbf{k}|\\
	& = & \dfrac{1}{\zeta}{\rm arcsin}(|\psi^{R}_{x(y)}|^{2}-|\psi^{L}_{x(y)}|^{2}).
	\label{zeta}
	\end{eqnarray}
	According to the slope distribution, we can reconstruct the phase of wave function
	\begin{equation}
	\varphi= \int \langle\hat{k}_{x}\rangle/|\textbf{k}| dx + \langle\hat{k}_{y}\rangle/|\textbf{k}| dy.
	\label{phase}
	\end{equation}
	The transverse wave function is,
	\begin{equation}
	|\psi\rangle=|\psi|e^{-i\varphi}
	\label{phi},
	\end{equation}
	where $|\psi|$ is the amplitude of wave function and can be obtained via projection measurement.

	In the experiment, after post-selected, the beam intensity $I^{R}_{x(y)}$ and $I^{L}_{x(y)}$ is detected by the ICCD camera. The probability in Eq. (\ref{zeta}) is $|\psi^{R(L)}_{x(y)}|^{2}=I^{R(L)}_{x(y)}/(I^{R}_{x(y)}+I^{L}_{x(y)})$. The distribution of wave function phase gradient is
	\begin{equation}
	\langle\hat{k}_{x(y)}\rangle/|\textbf{k}|=\dfrac{1}{\zeta}{\rm arcsin}(\dfrac{I^{R}_{x(y)}-I^{L}_{x(y)}}{I^{R}_{x(y)}+I^{L}_{x(y)}}),
	\end{equation}
	which is the result of Eq. (3) in main text.
	

  \section{II. Application: Simulation of wavefront sensing}
  
  A wavefront sensor is a device for measuring the aberrations of an optical wavefront. Here we show two kinds of simulated results via momentum weak measurement to reconstruct phase distributions with aberrations. 
  
  The parameters of simulation are the same as those of experiment, including wave length $\lambda=880\ nm$, spatial resolution $1024\times1024$ and coupling strength $\zeta=336$. In order to simulate the characteristics of photons detectors ICCD camera, Gaussian noise is applied to the intensity of light according to the dark count of ICCD camera and the fluctuation of photons.
  
  \subsection{Noise simulation}
  
  \begin{figure}[t]
  	\centering
  	\includegraphics[width=\linewidth]{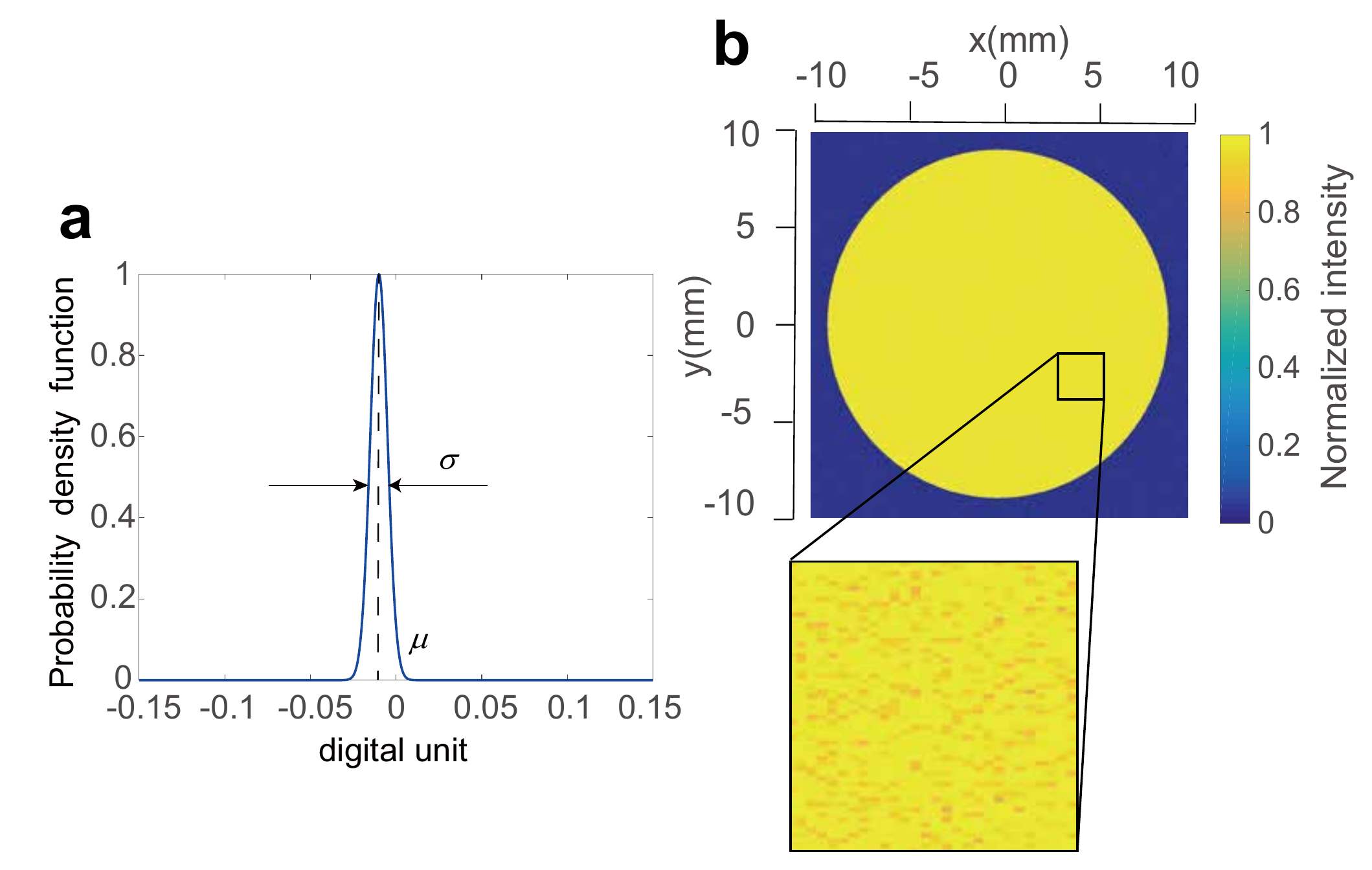}
  	\caption{ \textbf{Noise Simulation.} \textbf{a.} Normalized statistical representation of the noise distribution. \textbf{b.} Intensity distribution on the detector.}
  	\label{noise}
  \end{figure}
  In order to simulate the characteristics of experimental detectors, we applied Gaussian noise to the intensity of light. During the experiment, we found that the relative dark count of the ICCD is 0.01, and the relative fluctuation of  photons is 0.005. Therefore, the mean and variance of Gaussian noise are set to 0.01 and 0.005 respectively, which is shown in Fig. \ref{noise}a. The light intensity received by the detector is shown in Fig. \ref{noise}b.

  \subsection{Zernike aberration test}
  
		\begin{figure}[b]
			\centering
			\includegraphics[width=0.45\textwidth]{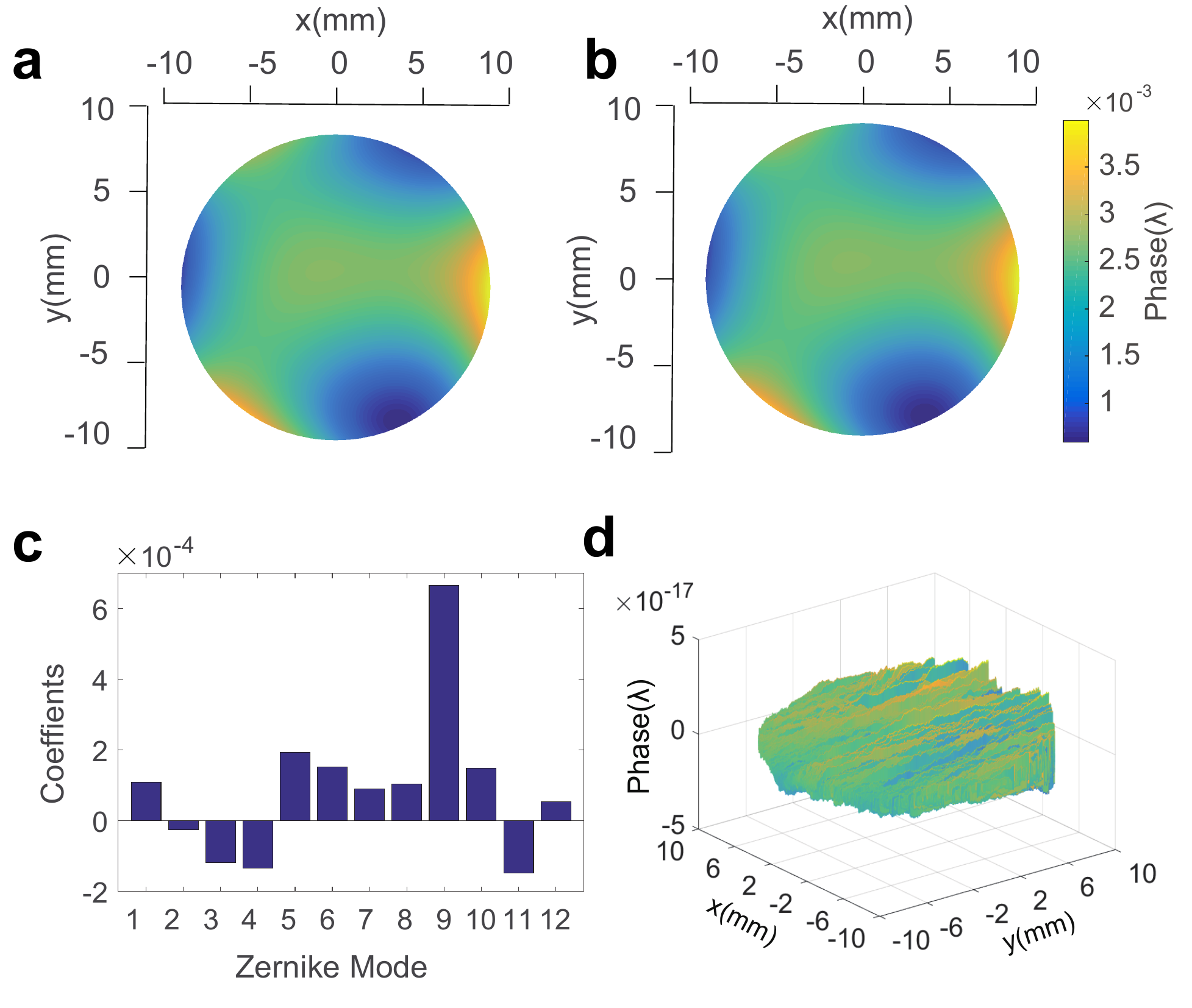}
			\caption{\label{Fig:1} \textbf{Numerically reconstructed wavefront based on weak measurement.} \textbf{a.} The original wavefront to be reconstructed ($RMS=0.0028\lambda, P-V=0.01\lambda$). \textbf{b.} Reconstructed wavefront. \textbf{c.}  Zernike coefficients of the wavefront. \textbf{d.} The residual wavefront error between \textbf{a.} and \textbf{b.} ($RMS=1.35\times10^{-17}\lambda$).}
		\end{figure}
		
The simulated results are shown in FIG. \ref{Fig:1}. The wavefront phase distribution contains 12 Zernike aberrations~\cite{zernike} (shown in FIG. \ref{Fig:1}c), which are randomly generated. Root mean square (RMS) and peak valley difference (P-V) of the original wavefront to be reconstructed (FIG. \ref{Fig:1}a) are $RMS=0.0028\lambda$ and $P-V=0.01\lambda$ respectively. In FIG. \ref{Fig:1}b, the reconstructed wavefront is obtained with the momentum weak value measurement. FIG. \ref{Fig:1}d shows the relative residual wavefront error, whose $RMS$ is $1.35\times10^{-17}\lambda$. 
The results show that, in simulation, momentum weak measurement is useful to reconstruct wavefront and detect the aberrations.

\subsection{Aberration with high spatial frequency detection }
\begin{figure}[t]
	\centering
	\includegraphics[width=0.45\textwidth]{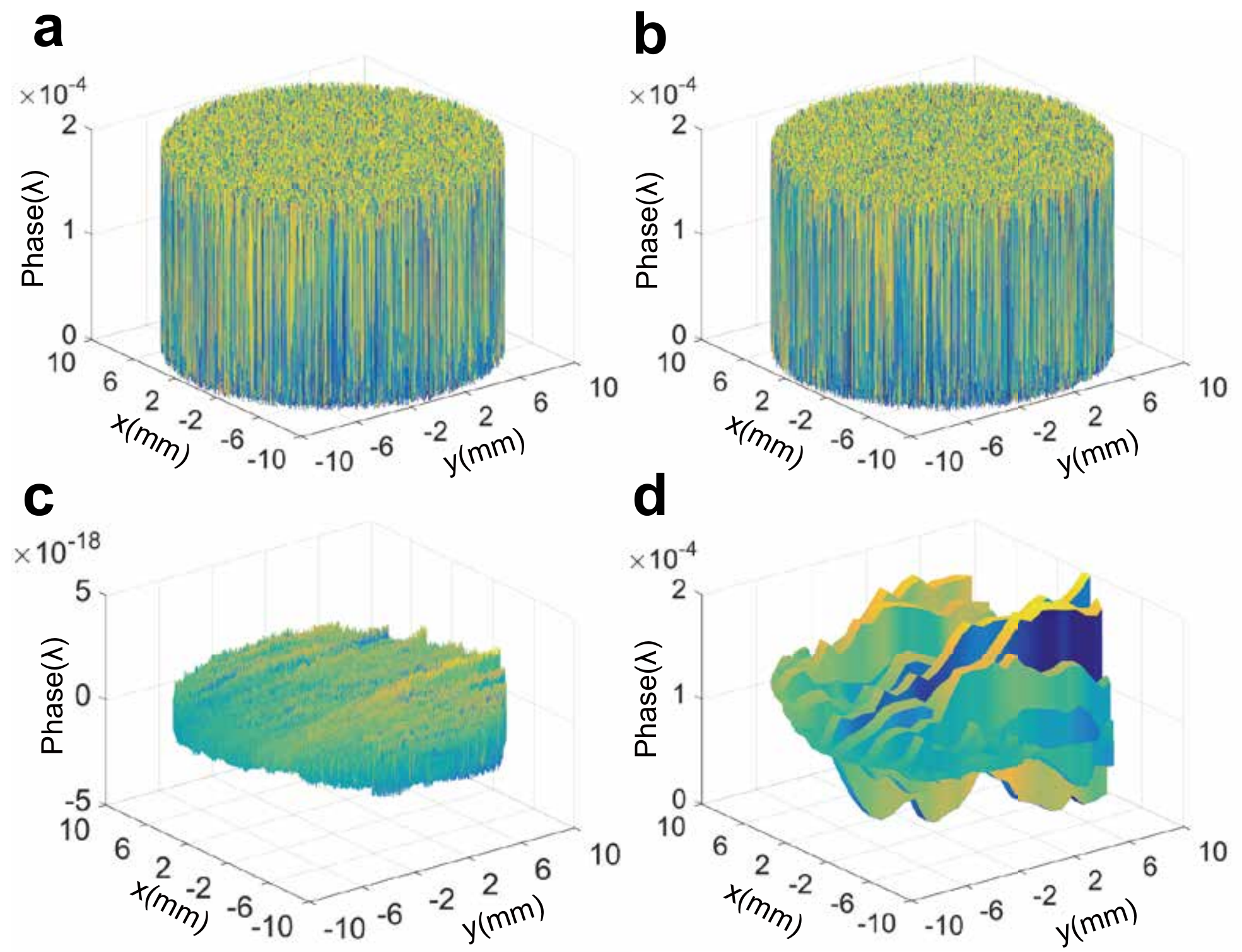}
	\caption{\label{Fig:2} \textbf{Numerically reconstructed wavefront with ultra-high spatial frequency.} \textbf{a.}  The ultra-high spatial frequency wavefront to be reconstructed ($RMS=0.0002\lambda, P-V=0.00015\lambda$). \textbf{b.} Reconstructed wavefront through weak measurement.  \textbf{c.} The residual wavefront error between \textbf{a} and \textbf{b} ($RMS=7.65\times10^{-19}\lambda$). \textbf{d.} Reconstructed wavefront through SHWFS with $25\times25$ microlens arrays.}
\end{figure}

Fig. \ref{Fig:2} shows the simulated results. The original wavefront with ultra-high spatial frequency aberrations is shown in FIG. \ref{Fig:2}a, in witch the $RMS=0.0002\lambda$, $P-V=0.00015\lambda$ and each pixel's phase is randomly generated. The reconstructed wavefront via momentum weak measurement and the relative residual error are shown in FIG. \ref{Fig:2}b and c, respectively. The RMS wavefront error is set to be $7.65\times10^{-19}\lambda$.  In contrast, the reconstructed wavefront by SHWFS (The number of microlens arrays is 25$\times$25) is shown in FIG. \ref{Fig:2}d. The RMS wavefront error is $6.87\times10^{-5}\lambda$, which means SHWFS can not reconstruct the wavefront with ultra-high spatial frequency.\\

\section{III. Zernike modes wavefront sensing algorithm}

\begin{figure}[t]
	\centering
	\includegraphics[width=\linewidth]{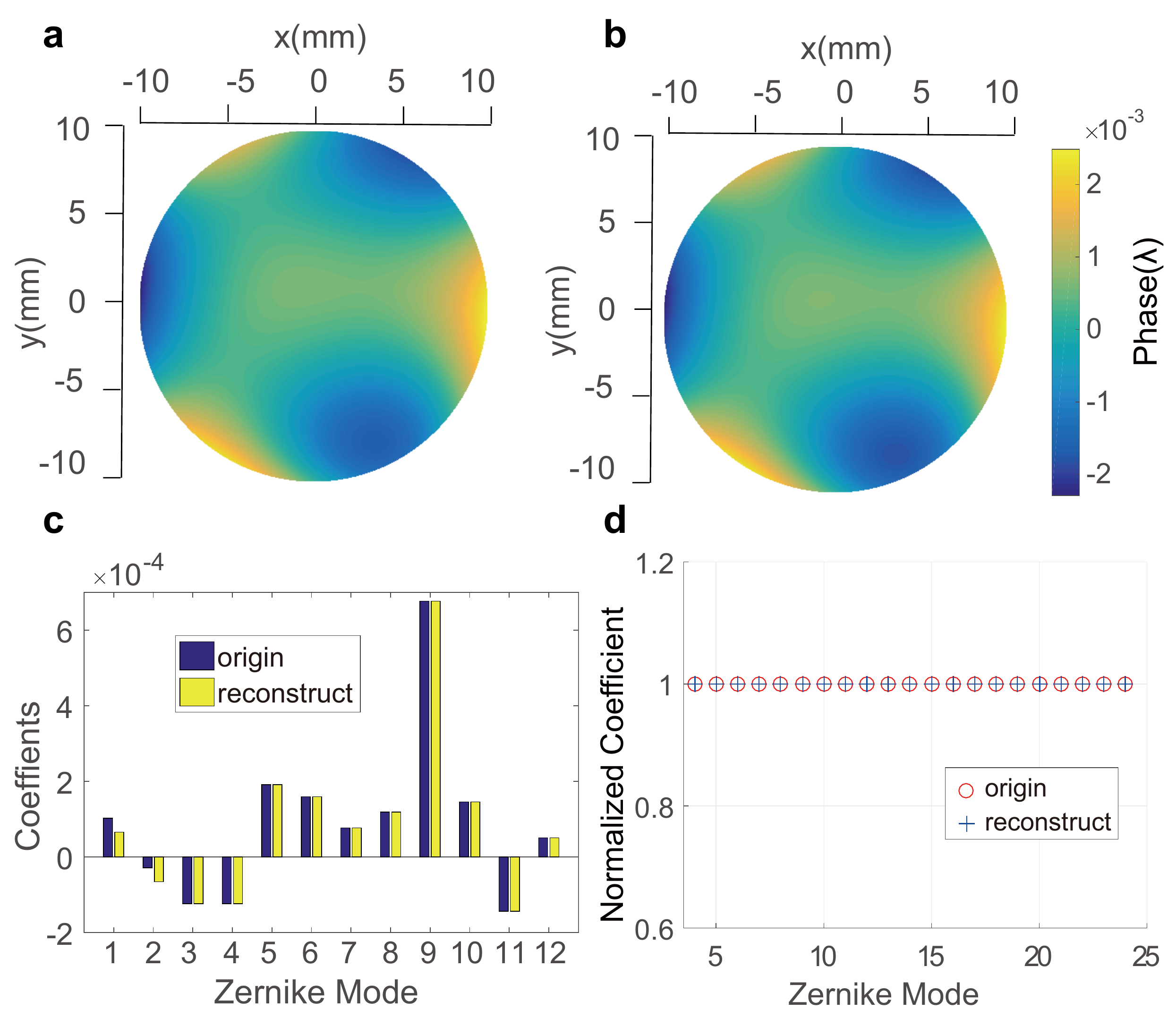}
	\caption{\textbf{Numerically reconstructed wavefront with multiorder Zernike modes.} \textbf{a.} The original wavefront to be reconstructed, \textbf{b.} The reconstructed wavefront. \textbf{c.} Zernike aberration restoration for each order Zernike mode. \textbf{d.} Zernike coefficients of the wavefronts in \textbf{a} and \textbf{b}.}
	\label{modelaw}
\end{figure}

Here we will introduce another wavefront recovery algorithm~\cite{lane1992wave} based on Zernike modes wavefront sensing. 
The wavefront can be expressed as a linear combination of two-dimensional orthogonal basis, for example, Zernike modes. Assuming the wavefront phase distribution is $\phi(x,y)$, which can be express as,
\begin{equation}
\phi(x,y)=\sum_{k=1}^{n}a_{k}Z_{k}(x,y),
\label{Zernike}
\end{equation}
where $a_{k}$ are coefficients of Zernike mode $Z_{k}(x,y)$. Differentiation on both sides of the Eq. (\ref{Zernike}), the relationship between wave front slope and Zernike mode coefficients can be obtained,
\begin{eqnarray}
\frac{\langle\hat{k}_{x}\rangle}{|\textbf{k}|}= \frac{\partial \phi}{\partial x} = \sum_{k=1}^{n}a_{k}\frac{\partial Z_{k}}{\partial x},\\
\frac{\langle\hat{k}_{y}\rangle}{|\textbf{k}|}= \frac{\partial \phi}{\partial x} = \sum_{k=1}^{n}a_{k}\frac{\partial Z_{k}}{\partial y}.
\end{eqnarray}
By projecting the measured wavefront slope information to the vector space composed of the partial derivatives of Zernike two-dimensional orthogonal basis $\partial Z_{k}/\partial x(y)$, the coefficients $a_{k}$ of wavefronts can be measured.

The simulation results are shown in Fig. \ref{modelaw}. Fig. \ref{modelaw}a. denotes the input wavefront, and Fig. \ref{modelaw}b. represents the wavefront reconstructed by the slope fitting method. The Zernike coefficients of Fig. \ref{modelaw}a. and Fig. \ref{modelaw}b. are shown in Fig. \ref{modelaw}c. 
In addition, we simulate the reconstructing ability of this method for each order Zernike mode, which is shown in \ref{modelaw}d. The input Zernike coefficient is consistent with the output.
The results show that this method can be used to restore the distorted wavefront with multi-order Zernike aberrations.
